\documentclass[aps,prd,twocolumn,superscriptaddress,nofootinbib]{revtex4-1}

\usepackage{amsmath}
\usepackage{mathptmx}
\usepackage{helvet}
\usepackage{courier}
\usepackage{mathrsfs}
\usepackage[dvips]{graphicx}

\begin{document}


\title{Constraining the strangeness content of the nucleon \\
by measuring the $\phi$ meson mass shift in nuclear matter}


\author{Philipp Gubler}
\email[]{pgubler@riken.jp}
\affiliation{RIKEN, Nishina Center, Hirosawa 2-1, Wako, Saitama, 351-0198, Japan}

\author{Keisuke Ohtani}
\email[]{ohtani.k@th.phys.titech.ac.jp}
\affiliation{Tokyo Institute of Technology, Meguro 2-12-1, Tokyo 152-8551, Japan}


\date{\today}

\begin{abstract}
The behavior of the $\phi$ meson at finite density is studied, 
making use of a QCD sum rule approach 
in combination with the maximum entropy method. 
It is demonstrated 
that a possible mass shift of the $\phi$ in nuclear matter is strongly correlated to the 
strangeness content of the nucleon, which is proportional to the 
strange sigma term, $\sigma_{sN} = m_s \langle  N | \overline{s}s | N \rangle$. 
Our results furthermore show that, depending on the value of $\sigma_{sN}$, the 
$\phi$ meson could receive both a positive or negative mass shift at nuclear matter 
density. 
We find that these results 
depend only weakly on 
potential modifications of the width of the $\phi$ peak and on assumptions made on the behavior 
of four-quark condensates at finite density. 
To check the stability 
of our findings, we take into account 
several higher order corrections to the operator product expansion, including $\alpha_s$-corrections, 
terms of higher order in the strange quark mass and terms of higher twist that have not been considered in earlier works. 
\end{abstract}


\maketitle

\section{\label{Intro} Introduction}
The strangeness content of the nucleon, $\langle  N | \overline{s}s | N \rangle$ 
is an important quantity, both for understanding the effects of 
strange quarks on the nucleon structure \cite{Ji} and 
the behavior of the strange quark condensate in dense matter. 
The second point is related to the fact that, via the Feynman-Hellmann 
theorem, the value of $\langle  N | \overline{s}s | N \rangle$ 
determines, to leading order in density $\rho$, 
the pace of restoration of chiral symmetry of the 
strange quark sector in nuclear matter: 
\begin{equation}
\langle \overline{s} s \rangle_{\rho} = \langle \overline{s} s \rangle_{0} + \langle  N | \overline{s}s | N \rangle \rho. 
\label{eq:chiralsym}
\end{equation}
The strange quark content of the nucleon also has implications that go beyond 
the physics of hadrons and the strong interaction, as it appears in the spin-independent 
elastic scattering cross sections of potential dark matter particles with nucleons 
and is one of the main sources of uncertainty for these cross sections \cite{Bottino,Ellis}, 
therefore strongly affecting experimental dark matter searches. 
Hence, it is crucial to determine this quantity with high precision. 

A variety of variables have been introduced in the literature to parametrize 
$\langle  N | \overline{s}s | N \rangle$. We will in this paper mainly use  
\begin{equation}
\sigma_{sN} = m_s \langle  N | \overline{s}s | N \rangle, 
\label{eq:strsigma}
\end{equation} 
and refer to it as the strange sigma term. This is a convenient quantity, 
because it is 
a renormalization group invariant, which is only related 
to the strange quark and does not directly depend on parameters related to 
$u$ and $d$ quarks, which have their own uncertainties.
\footnote{
Another commonly used variable is 
$y = \frac{\langle  N | \overline{s}s | N \rangle} {\langle  N | \overline{q}q | N \rangle} 
= \frac{2m_q}{m_s} \frac{\sigma_{sN}}{\sigma_{\pi N}}$, 
where $\langle  N | \overline{q}q | N \rangle$ and $m_q$ are averaged quantities over the $u$ and $d$ quarks. 
$\sigma_{\pi N}$ stands for the $\pi N$ sigma term and is defined as $\sigma_{\pi N} = 2m_q \langle  N | \overline{q}q | N \rangle$. 
Other parametrizations use 
$\sigma_0  = (1 - y)\sigma_{\pi N}$ or $f_{T_s} = \sigma_{sN}/m_{N}$, in which $m_N$ represents the nucleon mass.
}

In recent years, 
it has become possible to evaluate $\sigma_{sN}$ 
by direct lattice QCD calculations or by chirally extrapolating the available 
lattice data 
\cite{Young,Babich,Durr,Horsley,Bali,Semke,Freeman,Shanahan,Ohki,Alarcon,Engelhardt,Junnarkar,Jung,Gong,Alexandrou,Lutz,Ren}. 
These studies 
have shown that the value of $\sigma_{sN}$ is about a factor of 5 smaller than what was believed  
to be the correct range of values until about a decade ago. 
The values reported by the various groups however still show quite a large spread, lying roughly 
in the range of $0 \sim 70$ MeV, which indicates that these 
calculations still contain systematic uncertainties. 

We will in this paper discuss a different method for obtaining $\sigma_{sN}$, namely by experimentally 
measuring the mass of the $\phi$ meson in nuclear matter. 
The behavior of the $\phi$ meson 
as a function of density strongly depends on the value of $\sigma_{sN}$ and can therefore 
provide a strong constraint for this quantity once it is measured with sufficiently high precision. 
The relation between the $\phi$ meson mass and $\sigma_{sN}$ [shown in Fig. \ref{fig:sigmaN} and Eq.(\ref{eq:massshiftdep})] 
is derived from a recently developed method combining QCD sum rules with the 
maximum entropy method (MEM) \cite{Gubler}. 
This method allows us to extract the 
most probable spectral function 
directly from the sum rules, without having to make strong assumptions on its functional 
form. 
Moreover, by making use of MEM, it is possible to determine the mass shift of the $\phi$ meson quite 
precisely. As shown in Sec. \ref{MEMmock}, we have tested the reproducibility of the mass shift 
in a series of mock data tests, in which we have determined the precision of the mass shift 
extraction by MEM to be of the order of 5 MeV, which is good enough for our purposes.  

QCD sum rule studies of light vector mesons at finite density in fact already 
have quite a long history \cite{Hatsuda,Asakawa2,Koike,Jin,Hatsuda2,Klingl,Leupold2,Lee,Leupold3,Zschocke2,Kaempfer}. 
They have especially attracted 
much interest because QCD sum rules provide relations between 
the partial restoration of 
chiral symmetry in nuclear matter and modifications of meson spectra 
that could be measured in experiments \cite{Hayano,Leupold}. 
The early works on this subject have usually focused on the $\rho$ and 
$\omega$ channels and the relation between their mass shifts and 
chiral symmetry. By now, it is however understood that this 
relation is not a simple one, as the driving term for the modification of the spectrum entering into the sum rules 
contains not the most simple two-quark condensate but the more involved four-quark condensate. 
Moreover, for the $\rho$ meson, the spectral modification cannot be assumed to be a simple mass shift of the ground state peak, 
but is rather a combination of 
mass shift and broadening for which the sum rules only provide a relatively weak constraint \cite{Leupold2}. 

These issues are less severe for the $\phi$ meson. 
Due to the effects of the strange quark mass on the operator product expansion (OPE), 
it is for the $\phi$ meson channel mainly the dimension-four 
term that governs the modification of the spectral function and the dimension-six term (in which the four-quark 
condensate dominates) is merely a small correction with no large effects. 
As it has already been pointed out in earlier works \cite{Hatsuda,Klingl,Zschocke2}, 
this leads to an unambiguous relation between the $\phi$ meson mass shift and the 
strange sigma term, which appears at dimension four. 
Furthermore, even though the $\phi$ is expected to experience some broadening when put into 
nuclear matter, it is known that its width will not grow above the 100 MeV level and might even stay much below, 
namely around 75 MeV \cite{Polyanskiy} or at an even smaller value \cite{Muto}. 
Therefore, the $\phi$ meson will retain its character as a relatively narrow peak in the spectral 
function that facilitates its analysis as no severe complications from large broadening effects arise. 

Because of these advantages, one can expect that the predictive power of the sum rules 
of the $\phi$ is bigger than for the $\rho$ and $\omega$. To check whether this expectation is  
actually true and to quantify in what way the sum rules 
constrain the properties of the $\phi$ meson peak, and, thus, inversely, in what way 
the $\phi$ meson mass constrains the values of $\sigma_{sN}$, 
is the main goal of this paper. 

In Sec. \ref{Basics}, we will first recapitulate the basics of QCD sum rules and their 
application to finite density. 
Next, in Sec. \ref{OPEof}, the OPE of the $\phi$ meson channel in vacuum and at finite density 
will be given. 
Then, to demonstrate the ability of MEM to reproduce mass shifts at finite density, the results of 
our mock data analyses will be summarized in Sec. \ref{MEMmock}. 
Finally, the analysis results of the OPE data are provided and discussed in Secs. \ref{MEMOPE} and \ref{discussion}, while 
the conclusions are given in Sec. \ref{Conclusion}. 

\section{\label{Formalism} Formalism}
\subsection{\label{Basics} Basics of QCD sum rules}
As usual when working with QCD sum rules \cite{Shifman}, we start with the two-point function of 
an interpolating field coupling strongly to the $\phi$ meson: 
\begin{equation}
\Pi_{\mu\nu}(\omega,\vec{q}) = i\displaystyle \int dx^4 e^{iqx} \langle \mathrm{T} [j_{\mu}(x) j_{\nu}(0)] \rangle_{\rho}.
\label{eq:veccorr1}
\end{equation}
Here, the operator $j_{\mu}(x)$ is defined as $j_{\mu}(x) = \overline{s}(x) \gamma_{\mu} s(x)$, 
and $\langle \, \rangle_{\rho}$ stands for the expectation value with respect to 
the ground state of nuclear matter at $T=0$. Generally, $\Pi_{\mu\nu}(\omega,\vec{q})$ contains two 
independent Lorentz structures \cite{Gale}, but 
for the case of 
the $\phi$ meson at rest relative to the nuclear medium ($\vec{q} = 0$), there is 
only one such structure and it suffices to consider the contracted correlator defined as 
$\Pi(\omega^2) = -\frac{1}{3\omega^2} \Pi_{\mu}^{\mu}(\omega,\vec{q} = 0)$. 
From the analyticity of $\Pi(\omega^2)$, the 
dispersion relation 
\begin{equation}
\Pi(\omega^2) = \frac{1}{\pi}  \displaystyle \int_0^{\infty} ds \frac{\mathrm{Im}\Pi(s)}{s - \omega^2 -i\epsilon} 
\label{eq:displ}
\vspace{0.2cm}
\end{equation}
can be derived. 
The idea of the QCD sum rule approach is now to take $\omega^2$ as a large and negative number and to calculate the 
left-hand side of Eq.(\ref{eq:displ}) using the OPE. This results in a power series in $1/\omega^2$ with 
(Wilson) coefficients expressed as expansions in the strong coupling constant $\alpha_s$. On the right hand side, 
the function $\frac{1}{\pi} \mathrm{Im}\Pi(s)$ is viewed in terms of hadronic degrees of freedom 
that couple to the operator $j_{\mu}(x)$. 

Equation (\ref{eq:displ}) is in fact not yet the final form of the sum rule as the integral on its 
right-hand side is not convergent and thus a subtraction term is needed. The standard way to 
remedy this problem is the application of the Borel transform, which cancels any subtraction constant, 
renders the integral over $s$ convergent and furthermore improves the convergence of the OPE. 
This finally gives 
\begin{equation}
\Pi(M^2) = \frac{2}{M^2} \displaystyle \int^{\infty}_{0}d\omega
e^{-\omega^2/M^2} \omega A(\omega), 
\label{eq:Boreltrans2}
\end{equation}
where we have defined the spectral function $A(\omega)$ as $A(\omega) = \frac{1}{\pi} \mathrm{Im}\Pi(\omega^2)$. 
This is the final form of the sum rule that will be used in the analyses presented in this paper. 

\subsection{\label{OPEof} OPE of the $\phi$ meson channel in vacuum and at finite density}
The result of the OPE is generally obtained as shown below: 
\begin{equation}
\Pi_{\mathrm{OPE}}(M^2,\rho) = c_0(\rho) + \frac{c_2(\rho)}{M^2} + \frac{c_4(\rho)}{M^4} + \frac{c_6(\rho)}{M^6} +  \dots.  
\label{eq:ope1}
\end{equation}
In this work, we consider terms up to dimension six. 
The coefficients $c_i(\rho)$ have mass dimension $i$ and can contain 
logarithmic dependencies on the Borel mass $M$, which we 
have not explicitly spelled out in Eq.(\ref{eq:ope1}) for simplicity of notation. 

The coefficients in the vacuum ($\rho=0$) can be given as follows \cite{Shifman,Generalis,Chetyrkin,Loladze,Surguladze}: 
\begin{widetext} 
\begin{equation}
c_0(0) = \frac{1}{4 \pi^2} \Big( 1 + \frac{\alpha_s}{\pi} \Big), \hspace{1cm}
c_2(0) = \frac{m_s^2}{4 \pi^2} \Biggl[-6 - 4\frac{\alpha_s}{\pi} \bigg(4 - 6\log \Big(\frac{M}{\mu} \Big) + 3\gamma_E   \bigg)  \Biggr], \label{eq:operesult1} 
\end{equation}
\begin{align}
c_4(0) =& \frac{1}{12}\Big(1 + \frac{7}{6}\frac{\alpha_s}{\pi} \Big) \Big \langle \frac{\alpha_s}{\pi} G^2 \Big \rangle 
+ 2m_s \Big(1 + \frac{1}{3}\frac{\alpha_s}{\pi} \Big)\langle \bar{s} s \rangle + \frac{3}{4\pi^2}m_s^4 \bigg[1 + 4 \log\Big(\frac{M}{\mu} \Big) -2\gamma_E    \bigg] \nonumber \\
&-\frac{1}{6\pi^2}m_s^4 \frac{\alpha_s}{\pi} \Biggl[35 - 3\pi^2 - 24 \zeta(3) + 3 \bigg(2\log \Big(\frac{M}{\mu} \Big) - \gamma_E   \bigg) +  18\bigg(2\log \Big(\frac{M}{\mu} \Big) - \gamma_E   \bigg)^2 \Biggr], \label{eq:operesult2} \\
c_6(0) =& -\frac{112}{81} \pi \alpha_s \kappa_0 \langle \bar{s} s \rangle^2 + \frac{1}{18} m^2_s \Big \langle \frac{\alpha_s}{\pi} G^2 \Big \rangle -\frac{4}{3} m^3_s \langle \bar{s} s \rangle. 
\label{eq:operesult3}
\end{align}
\end{widetext}
Up to dimension four, we have taken into account the first order $\alpha_s$ corrections to all the Wilson coefficients, 
which were calculated already a long time ago \cite{Chetyrkin,Loladze}, but have to our 
knowledge not been taken into account in the available sum rule studies on the $\phi$ meson. 
These corrections, however, turn out to be quite small, namely around $10\,\%$ of the leading 
order terms or smaller, with the exception of the $m_s^2$ and $m_s^4$ terms. 
Note that for full consistency, we, in principle, should have included the terms of $\alpha_s^2$ in the 
perturbative dimension-zero term of Eq.(\ref{eq:operesult1}), since they correspond to the 
hard parts of the diagrams that give the first order $\alpha_s$ corrections to the Wilson 
coefficients of the gluon condensate, which we have taken into account in Eq.(\ref{eq:operesult2}). 
As these $\alpha_s^2$ terms, in contrast to their counterparts appearing with the gluon condensate, however do not 
give any contribution to the density dependence of the OPE and hence to the $\phi$ meson mass shift, we have 
ignored them here. 

\begin{figure}
\includegraphics[width=7.5cm]{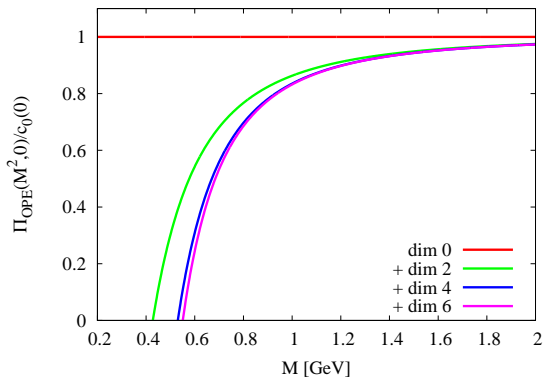}
\caption{\label{fig:OPE.plot} The OPE terms in vacuum, given in Eqs.(\ref{eq:operesult1}-\ref{eq:operesult3}), as a function of the Borel mass $M$. The contributions are shown relative 
to the leading order OPE term, $c_0(0)$.}
\end{figure}
Furthermore, we include several higher order terms in the strange quark mass $m_s$, which 
have not been considered before: $m_s^4$ \cite{Chetyrkin}, 
$m^2_s \big \langle \frac{\alpha_s}{\pi} G^2 \big \rangle$ and $m^3_s \langle \bar{s} s \rangle$ \cite{Generalis}. 
These corrections have also turned out to be small compared to other terms of the same dimension. 
Note that there are in fact more condensates of dimension six, that can, in principle, appear in the OPE. Specifically, these are 
$m_s \langle \bar{s} g \sigma \cdot G s\rangle$ and $\big \langle g^3 G^3 \big \rangle$, whose Wilson coefficients, 
however, are known to vanish at leading order in $\alpha_s$ \cite{Generalis}. 
One further point worth mentioning here is concerned with the four-quark condensate term at dimension six. To obtain its form given 
in Eq.(\ref{eq:operesult3}) we have assumed that the vacuum saturation approximation holds, and have 
parametrized the possible breaking of this approximation by the parameter $\kappa_0$. In vacuum we will 
assume that $\kappa_0$ is 1 (and thus that the vacuum saturation approximation is exact), but will consider 
its deviation from 1 later in the finite density case. 

To get an idea of the behavior of the various terms of Eqs.(\ref{eq:operesult1}-\ref{eq:operesult3}), they are plotted in Fig. \ref{fig:OPE.plot}. 
For drawing this plot, we have used 
$\alpha_s=0.5$ \cite{Bethke}, $\big \langle \frac{\alpha_s}{\pi} G^2 \big \rangle=0.012\pm0.0036\,\mathrm{GeV}^4$ \cite{Colangelo}, 
$m_s=128\pm7\,\mathrm{MeV}$ \cite{Beringer}, $\langle \bar{s} s \rangle=(0.6\pm0.1)\langle \bar{q} q \rangle$ \cite{Dominguez}, 
$\langle \bar{q} q \rangle = -(0.232\pm0.06)^3\,\mathrm{GeV}^3$ \cite{Aoki}, and $\kappa_0=1$. 
All these values are given at renormalization scale $1\,\mathrm{GeV}$. 

It is observed in Fig. \ref{fig:OPE.plot} that the qualitative properties of the OPE are essentially determined by the 
first three terms while the fourth term, which is dominated by the four-quark condensate, is only a small correction. This is fortunate, as the four-quark condensate 
is not well known, which in the parametrization used here translates into our lack of knowledge of the actual value of $\kappa_0$. 

Next, for investigating the finite density case, we need to calculate 
the $\rho$ dependence of the coefficients $c_i(\rho)$ in Eq.(\ref{eq:ope1}). 
To obtain $c_i(\rho)$ for general values of $\rho$ is a very difficult task that is beyond our ability 
at the present time. 
What we, however, can do, is to restrict ourselves to low densities and assume that the linear density approximation 
is valid for the density regime that we are interested in. We are in this work mainly interested in the modification 
of the $\phi$ meson at nuclear matter density and there is evidence that this approximation indeed 
works well there \cite{Hatsuda,Hatsuda2}. We will further discuss this point in Sec. \ref{discussion} 
and for the moment just assume that we are considering densities at which the above assumption 
is valid. 
The $\rho$ dependence of the coefficients $c_i(\rho)$ at linear order has already been discussed many 
times in the literature \cite{Hatsuda,Klingl,Leupold3}; the main focus, however, was usually laid on the $\rho$ and 
$\omega$ channels. We here give the result for the $\phi$ channel, which again includes several new terms 
that have not been taken into account in the works published so far: 
\begin{widetext} 
\begin{align} 
c_0(\rho) =& c_0(0), \hspace{1cm}
c_2(\rho) = c_2(0), \label{eq:finitedensity1} 
\end{align}
\begin{align} 
c_4(\rho) =& c_4(0) + \rho \Biggl[ -\frac{2}{27} \Big(1 + \frac{7}{6}\frac{\alpha_s}{\pi} \Big)M_{N} 
+ \frac{56}{27}m_s \Big(1 + \frac{61}{168}\frac{\alpha_s}{\pi} \Big) \langle N|\bar{s} s| N \rangle \nonumber \\
&+ \frac{4}{27}m_q \Big(1 + \frac{7}{6}\frac{\alpha_s}{\pi} \Big) \langle N|\bar{q} q| N \rangle
+ \Big(1 - \frac{5}{9}\frac{\alpha_s}{\pi} \Big)A^s_2 M_N - \frac{7}{12}\frac{\alpha_s}{\pi} A^g_2 M_N \Biggr], \label{eq:finitedensity2} \\
c_6(\rho) =& c_6(0) + \rho \Bigl[ -\frac{224}{81} \pi \alpha_s  \kappa_{N} \langle \bar{s} s \rangle \langle N|\bar{s} s| N \rangle
-\frac{104}{81} m_s^3  \langle N|\bar{s} s| N \rangle \nonumber \\
&+\frac{8}{81} m_s^2 m_q \langle N|\bar{q} q| N \rangle
-\frac{4}{81} m_s^2 M_N
-\frac{3}{4} m_s^2 A^s_2 M_N - \frac{5}{6} A^s_4 M_N^3
 \Bigr]. \label{eq:finitedensity3}
\end{align}
\end{widetext} 
The novel terms here are the $\alpha_s$ corrections of the terms at dimension four, a term related to 
a twist-2 gluonic operator at dimension four and several terms of higher order in $m_s$ at dimension six. 
Most of these terms only have a small effect and do not much change the earlier results. The only 
exception is the twist-2 gluonic operator at dimension four, which is proportional to the first moment of the gluonic parton distribution 
$A^g_2$. This term is in fact almost as large as the twist-2 strange quark operator (proportional to $A^s_2$), and therefore 
cannot be ignored. The definitions of $A^s_2$,  $A^s_4$ and $A^g_2$, which are all moments over parton 
distribution functions, can be found for instance in \cite{Jin2}. 

For the numerical evaluation of Eqs.(\ref{eq:finitedensity1}-\ref{eq:finitedensity3}), we use $M_N=940\,\mathrm{MeV}$, $2m_q\langle N|\bar{q} q| N \rangle = 45\pm7$ MeV \cite{Gasser}, 
$A^s_2=0.044 \pm 0.011$, 
$A^s_4=0.0011\pm0.0004$, and $A^g_2=0.359\pm0.146$. The last three values have been extracted numerically from the parton distributions given in \cite{Martin}. 
The breaking of the factorization assumption of the four-quark condensates is parametrized using $\kappa_{N}$. Here, we follow the 
treatment of \cite{Zschocke2} and take 
into account the possibility that $\kappa_{N}$ can differ from 
the vacuum value $\kappa_{0}$. Specifically, we will consider the range $\kappa_N=1\pm1$. 
Note that we have not explicitly included twist-4 terms, which, in principle, can appear at dimension six. 
In \cite{Hatsuda2,Choi}, 
the magnitude of these terms has been estimated for the $\rho$ ($\omega$) meson case to be 1.36 (2.29) times 
the corresponding twist-2 contribution of the same dimension. 
For the calculation of this paper, we assume that the ratio between twist-2 and twist-4 terms has the same order 
of magnitude for the $\phi$ meson case and take  the average value of 1.83 for this ratio. It is clear that 
this only a very crude estimate and we therefore will attach to it an uncertainty of $100\,\%$ 
and use $1.83\pm1.83$ 
when computing the error of the final results. 
Our poor knowledge of the twist-4 contribution does, however, not cause serious problems, 
as its contribution is negligibly small. 
Increasing for instance the above ratio by a factor of 2 only results 
in a mass shift of the $\phi$ meson peak of at most 1 MeV at 
nuclear matter density. 
As explained in the introduction, we do not assume any value for the strange sigma term $m_s \langle N|\bar{s} s| N \rangle$, but treat it 
as a free parameter. 

Let us here also make a few comments on the derivation of Eq.(\ref{eq:finitedensity2}). 
One might wonder where the term containing the (light quark) sigma term $2m_q\langle N|\bar{q} q| N \rangle$ comes from, even though we are 
considering a correlator of an interpolating field constructed from only strange quarks. This term arises from the density dependence of the 
gluon condensate, which to leading order in $\rho$ is proportional to the nucleon mass in the chiral limit \cite{Drukarev,Cohen}: 
$\big \langle \frac{\alpha_s}{\pi} G^2 \big \rangle_{\rho} = 
\big \langle \frac{\alpha_s}{\pi} G^2 \big \rangle_{\rho=0} - \frac{8}{9}(M_N - \sigma_{\pi N} - \sigma_{sN}) \rho$. 

Furthermore, it is also worth mentioning that the derivation of the twist-2 gluonic operator term is somewhat nontrivial as it 
can only be systematically obtained by taking into account the concept on non-normal ordered operators and their expectation values 
with respect to the nuclear matter ground state, as only for these the OPE is generally well defined in the chiral limit \cite{Chetyrkin2}. 
This procedure corresponds to subtracting out the soft contributions of perturbative quark propagators with attached gluon lines. 
A detailed account of how the related actual calculations can be done has been given recently in \cite{Zschocke} for the case of 
heavy-light quark pseudoscalar mesons and we here have followed the method put forward in that publication to 
obtain this twist-2 gluonic operator term of dimension four. For details, we thus 
refer the reader to \cite{Zschocke} and the references cited therein. 
\footnote{Let us however state here that we could not reproduce one small part of the formulas given in 
\cite{Zschocke}. Specifically, we got $\log\frac{\mu^2}{m_q^2} - \frac{1}{2}$ instead of 
$\log\frac{\mu^2}{m_q^2} - \frac{1}{3}$ in a factor appearing in the last term of Eq.(56) of \cite{Zschocke}. 
We will in this study use our own result, but note that this disagreement only has a very small numerical effect on 
the OPE.}

\section{Analysis results of the sum rules in vacuum and finite density}
\subsection{\label{MEMmock} MEM test analysis of mock data}
\label{mockdata}
Before directly analyzing the OPE of the previous section, we here at first will discuss the results 
of some test analysis of artificial mock data. This sort of test is important for confirming what properties of the spectral 
function MEM is able to reproduce and thus for getting an idea on the systematic error of this approach.
For details of MEM, we refer the reader to \cite{Jarrel,Asakawa}, and for practical details specific to the application of 
MEM to QCD sum rules, to \cite{Gubler,Gubler2}. 

\begin{figure}
\includegraphics[width=7.5cm]{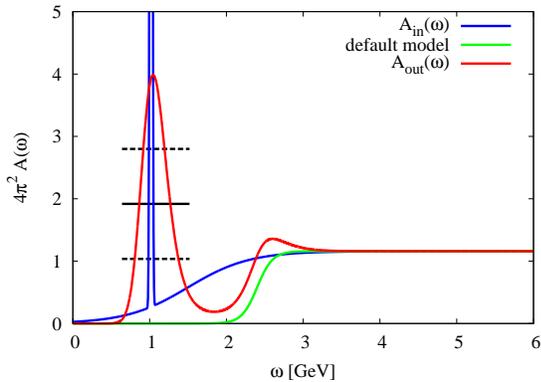}
\caption{Result of the MEM analysis of mock data ($A_{\mathrm{out}}(\omega)$) is plotted as a red line. 
The mock data were constructed 
from $A_{\mathrm{in}}(\omega)$, which is shown by the blue line.}
\label{fig:mock}
\end{figure}

First, we show the result of the mock data analysis of a spectral function resembling the one 
of the $\phi$ meson channel in the vacuum. For this, we use a realistic input spectral function, which has been fitted to experimental 
data in \cite{Shuryak},  
create mock OPE data by substituting this spectral function into the right-hand side of Eq.(\ref{eq:Boreltrans2}) 
and then analyze these mock data with MEM. 
The analyzed Borel mass region is taken to be the same as for the 
OPE data analysis of the following subsection. Specifically, we take $M_{\mathrm{min}} = 0.56\,\mathrm{GeV}$ 
for the lower boundary, which is determined from the convergence criterion of the OPE, 
which demands that the highest order OPE term should be smaller than $10\,\%$ of the leading perturbative 
term of dimension zero. For the upper boundary, we use $M_{\mathrm{max}} = 1.10\,\mathrm{GeV}$. This rather small 
value for $M_{\mathrm{max}}$ is taken to suppress the continuum contributions to the sum rules. 
The result $A_{\mathrm{out}}(\omega)$ is plotted in 
Fig. \ref{fig:mock}, in which the input spectral function $A_{\mathrm{in}}(\omega)$ 
and the employed default model (which is an input of the MEM analysis \cite{Jarrel,Asakawa}) 
are also shown for comparison. 
We evaluate the peak position by taking 
the average of $\omega$ over the peak region, which 
is defined as the interval for which the spectral function 
takes values above half of the maximum peak height. 
We will use the same prescription for the rest of this 
paper. 
This then leads to a value lying 
$44\,\mathrm{MeV}$ above the input peak position, which 
gives an idea of the quantitative accuracy of MEM for this particular 
quantity. 
We will, however, see below that when considering mass 
shifts, the precision will be considerably improved. 
It is furthermore understood from Fig. \ref{fig:mock} that the output $\phi$ peak is strongly broadened due 
to artificial MEM effects, which indicates that obtaining meaningful information on the width of the $\phi$ meson at finite density 
from our MEM approach will be a rather difficult task. We thus in the following concentrate our 
efforts only on the mass shift. 

\begin{figure}
\includegraphics[width=7.5cm]{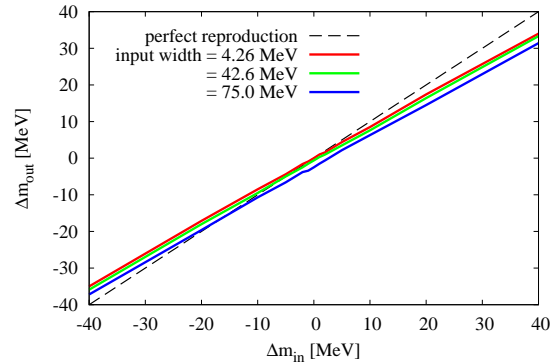}
\caption{Result of the MEM analysis of mock data with mass shifted peaks and various degrees of broadening. The 
mass shift obtained by the analysis is given as a function of the input mass shift of the mock data.}
\label{fig:repro}
\end{figure}

To test the quantitative ability of MEM to reproduce such a mass shift and how a possible broadening of 
the peak will influence the result, we will as a next step study several kinds of mock data, which include 
mass shifted peaks with various degrees of broadening. 
A few representative results are shown in Fig. \ref{fig:repro}, where input mass shifts are compared with 
their MEM outputs. 
As one can observe from this plot, the mass shifts 
are mostly reproduced well, the error only being a few MeV. 
It is however also seen that the mass shift is underestimated especially 
for large positive mass shifts. 
We will take this effect into account when extracting 
the $\phi$ meson mass shift from the actual OPE data. 
We furthermore note 
that broadening somewhat decreases the MEM output mass, 
which introduces some uncertainty into our results. 

We have also investigated the effect of a possible dependence of the continuum on the density. 
Even though we do not have any detailed knowledge on the behavior of the continuum in dense matter, 
we can make a crude estimate by assuming that it is dominated by freely propagating kaons with 
properties modified by the background density. 
It is known from effective models based on chiral symmetry that the kaons on the average 
receive a negative mass shift of below 50 MeV at nuclear matter density \cite{Waas,Tolos}, 
which can be translated into a shift (of below 100 MeV) of the continuum towards smaller energies. 
We have also considered the possibility of the continuum becoming smoothed out due to density effects and have 
allowed the gradient of the continuum to decrease up to 10 \%. 
Including these modifications into our analysis has again an effect of at most a few MeV on the peak position 
of the $\phi$ meson and hence leads to a further systematic uncertainty. 
Let us furthermore stress here that the assumptions on the behavior of the continuum mentioned above 
are only used to generate mock data for the MEM test analysis presented here. 
In the MEM analysis of the real OPE data of the next section, no assumptions are made on the actual 
behavior of the continuum.  
 
All the effects discussed in the last two paragraphs will be taken into account when evaluating the error of 
our final results. 

\subsection{\label{MEMOPE} MEM analysis of OPE data in vacuum and at finite density}
After the investigation of mock data of the last section, we are now in a position to 
study the actual OPE and to give an accurate interpretation of the obtained results. 

Let us start with the spectral function in vacuum, for which we analyze 
the OPE data of Eqs.(\ref{eq:operesult1}-\ref{eq:operesult3}). The result closely resembles 
the one of Fig. \ref{fig:mock} and we thus do not show it here. For the peak position ($m_{\phi}$), 
we get a value of 1.075 GeV, which lies 56 MeV above the experimental value 
of 1.019 GeV. Note that we have deliberately chosen a rather small value for the 
strange quark condensate to get this mass. This is done in purpose of starting the analysis 
from a spectral function in the vacuum that is as realistic as possible, 
as higher quark condensate values would lead to an even larger $m_{\phi}$. 

\begin{figure}
\includegraphics[width=7.5cm]{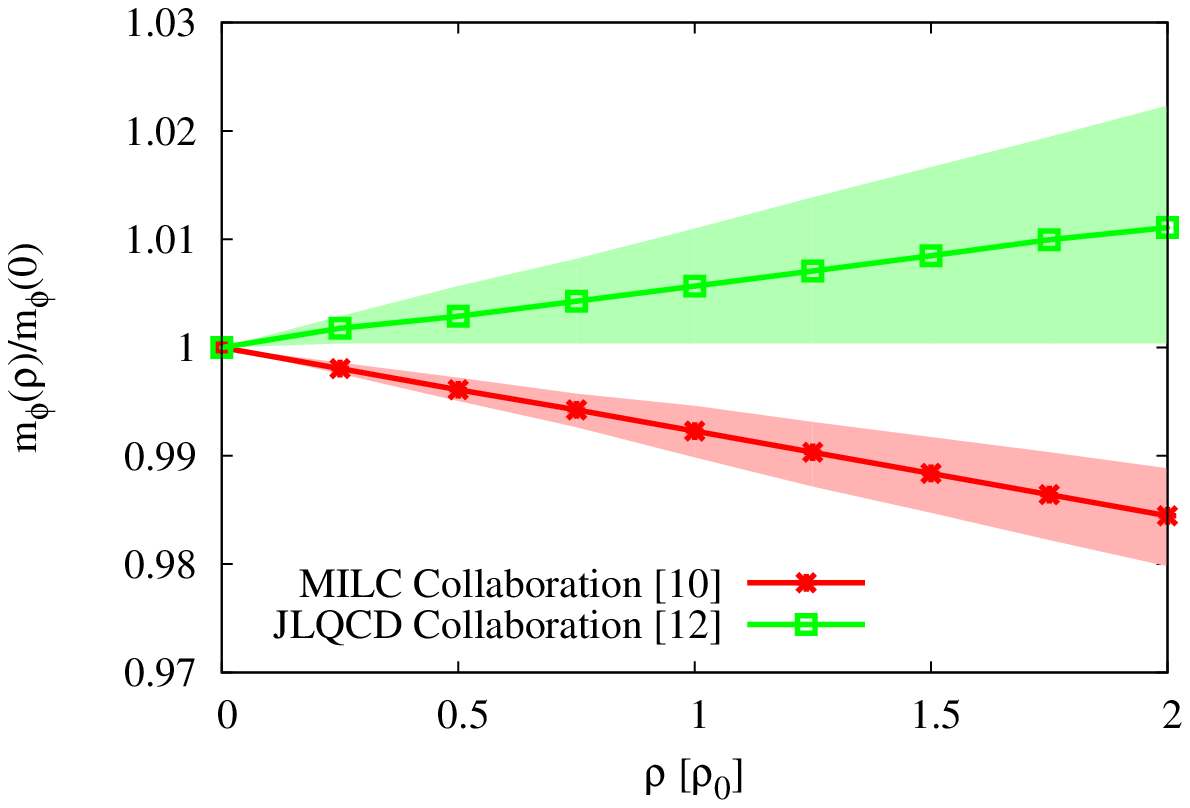}
\includegraphics[width=7.5cm]{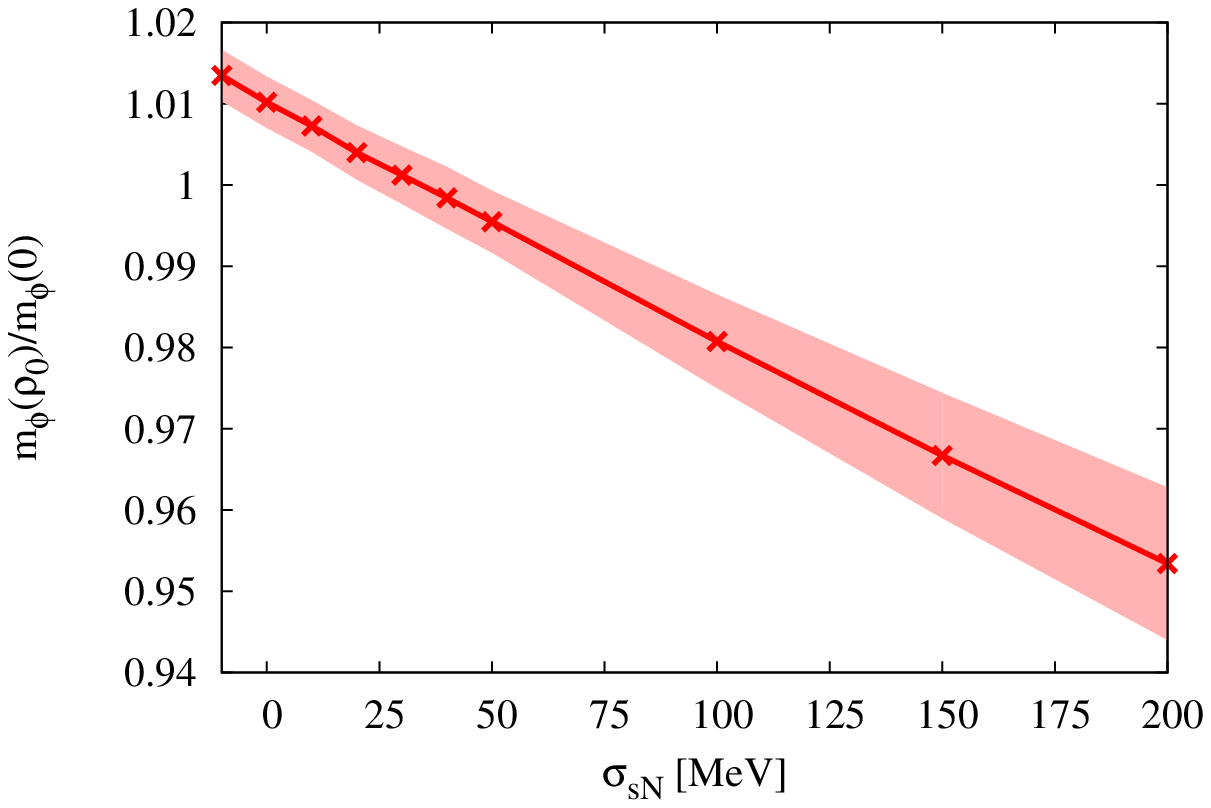}
\caption{(Upper plot) Peak position of $\phi$ meson 
as a function of the density $\rho$, 
for value ranges of the strange sigma term $\sigma_{sN}$, 
obtained from the MILC \cite{Freeman} and JLQCD \cite{Ohki} lattice QCD collaborations. 
The $\sigma_{sN}$ values are $61 \pm 9$ MeV for MILC and $8 \pm 21$ MeV for JLQCD. 
(Lower plot) Peak positions of the $\phi$ meson at nuclear matter density $\rho_0$ as a function of 
$\sigma_{Ns} = m_s \langle N|\bar{s} s| N \rangle$. 
For both plots, 
the peak positions are given relative to the $\phi$ mass in vacuum.}
\label{fig:sigmaN}
\end{figure}

Next, we proceed to the main subject of interest of this paper, the behavior of the $\phi$ meson 
at finite density. As a first example, we choose two values of the strange sigma term,  
provided by recent lattice QCD calculations \cite{Freeman,Ohki}, for which we 
have intentionally chosen results that lie on the lower 
and upper range of the values reported during the past few years. 
They will therefore provide a lower  
and upper limit for the mass shift of the $\phi$ meson, based on these lattice results. 
The behavior of the $\phi$ meson mass as a function of density is shown in the 
upper plot of Fig. \ref{fig:sigmaN}, where it is seen that the 
$\phi$ meson mass shift at nuclear matter density lies roughly in the range of 
$+10\,\mathrm{MeV}$ $\sim$ $-10\,\mathrm{MeV}$. 

This result is especially interesting in view of the fact that earlier sum rule studies have all 
\cite{Hatsuda,Asakawa2,Koike,Jin,Klingl,Zschocke2,Kaempfer} 
obtained a 
negative mass shift at 
nuclear matter density, while here we get both the possibility of a positive and negative mass shift, 
depending on the value of $\sigma_{sN}$. The reason for this discrepancy is twofold. First, the recent 
lattice QCD 
values of the strange sigma term are much 
smaller than those that had been used until about a decade ago, which significantly reduces the contribution 
of this term to the OPE of Eq.(\ref{eq:finitedensity3}). Furthermore, the twist-2 gluonic 
term of dimension four, which was not considered in these works, has turned out to have 
quite a large effect, leading to a further increase of the mass. 
Hence, the situation is now quite different from what it used to be and it is 
at present not even clear whether there will be a positive, negative or any mass shift at all 
at nuclear matter density. 

In this context, let us mention the works using hadronic models with 
phenomenologically determined effective Lagrangians \cite{Klingl,Klingl2,Oset,Cabrera}, 
which at normal nuclear matter density get a small but negative mass shift of 
below $10\,\mathrm{MeV}$ and a width about 
an order of magnitude larger than the vacuum value. 
As can be observed in Fig. \ref{fig:sigmaN}, this is consistent with our QCD sum rule result 
and some of the recent lattice QCD computation ranges of $\sigma_{sN}$, but would 
exclude too-small values of the strange sigma term, for which the mass shift is positive. 

As explained in the introduction, we do not choose any specific value of the 
strange sigma term, but study the modification of the $\phi$ meson more generally as a 
function of  this parameter. The result of this investigation is given in the lower plot of 
Fig. \ref{fig:sigmaN}, where 
the $\phi$ meson mass at nuclear matter density is shown as a function of 
$\sigma_{sN}$. Here, the error band includes the uncertainties of 
$A^s_2$, $A^s_4$, $A^g_2$, $2m_q\langle N|\bar{q} q| N \rangle$, $\kappa_{N}$, and of  
the twist-4 terms of dimension six. 
Furthermore, the systematic errors of the MEM analysis 
stemming from the possible broadening of the $\phi$ meson peak and the modification of the continuum, discussed at the end of Sec. \ref{mockdata}, 
are also taken into account. 
Figure \ref{fig:sigmaN} clearly demonstrates that there is an (almost) linear relationship between the 
$\phi$ meson mass shift and $\sigma_{sN}$. Altogether, the result of Fig. \ref{fig:sigmaN} can most simply be fitted by 
a constant plus a term linear in $\sigma_{sN}$: 
 \begin{equation}
\frac{m_{\phi}(\rho)}{m_{\phi}(0)} - 1  = \Biggl[b_0 - b_1\Bigl(\frac{\sigma_{sN}}{1\,\mathrm{MeV}}\Bigr) \Biggr] \frac{\rho}{\rho_0}, 
\label{eq:massshiftdep}
\end{equation}
$\rho_0$ representing the normal nuclear matter density. 
Our fit gives $b_0=(1.00 \pm 0.34) \cdot 10^{-2}$ and $b_1=(2.86 \pm 0.48) \cdot 10^{-4}$, which means that 
the mass shift changes its sign at a $\sigma_{sN}/1\,\mathrm{MeV}$  value of $b_0/b_1 = 34.9\pm13.1$. 
Using the variable $y$ instead of $\sigma_{sN}$, we get $0.174\pm 0.040$ for the slope parameter (which 
corresponds to $C/y$ in \cite{Hatsuda}, where a value of $0.15\pm0.045$ was obtained) 
with the sign of the mass shift switching at $y=(5.74\pm2.34) \cdot 10^{-2}$. 

\section{\label{discussion} Discussion}
Let us try to understand the result of Fig. \ref{fig:sigmaN} by looking at the OPE of 
Eqs.(\ref{eq:finitedensity1}-\ref{eq:finitedensity3}) a bit more closely. From our discussion of the OPE in vacuum [Eqs.(\ref{eq:operesult1}-\ref{eq:operesult3}], see 
also Fig. \ref{fig:OPE.plot}), we know that the properties of the $\phi$ meson are essentially determined by 
the OPE terms up to dimension four. As the dimension-zero and -two terms do not have any density dependence, 
one can therefore expect that the finite density contributions to the dimension-four terms will control the modification 
of the $\phi$ peak. These terms of linear order in $\rho$ have the following form: 
\begin{equation}
\begin{split}
 &-\frac{2}{27} \Big(1 + \frac{7}{6}\frac{\alpha_s}{\pi} \Big)M_{N} 
+ \frac{56}{27}m_s \Big(1 + \frac{61}{168}\frac{\alpha_s}{\pi} \Big) \langle N|\bar{s} s| N \rangle \\
&+ \frac{4}{27}m_q \Big(1 + \frac{7}{6}\frac{\alpha_s}{\pi} \Big) \langle N|\bar{q} q| N \rangle
+ \Big(1 - \frac{5}{9}\frac{\alpha_s}{\pi} \Big) A^s_2 M_N \\
&- \frac{7}{12}\frac{\alpha_s}{\pi} A^g_2 M_N. 
\end{split}
\label{eq:dim4}
\end{equation}
Written down numerically, this gives:
 \begin{equation}
\begin{split}
 &-82.5\,\mathrm{MeV} 
+ 2.19 \sigma_{sN} \\
&+ 3.95\,\mathrm{MeV}
+ 37.7\,\mathrm{MeV} - 31.3\,\mathrm{MeV} \\
&= 2.19 \Biggl[\Bigl(\frac{\sigma_{sN}}{1\,\mathrm{MeV}}\Bigr) - 32.9\Biggr]\,\mathrm{MeV}.
\end{split}
\label{eq:dim4.1}
\end{equation}
Here, a positive coefficient results in a negative mass shift and vice versa. Therefore it is easily understood 
from the numbers above that for small values of $\sigma_{sN}$ a positive mass shift is observed, which 
turns into a negative one at $\sigma_{sN} = 32.9\,\mathrm{MeV}$. 
This quite accurately describes the situation observed in Fig. \ref{fig:sigmaN} and shows that it is 
indeed the dimension-four terms that almost completely determine the $\phi$ meson mass shift 
at finite density.  

Another important point that needs some discussion is how our results can be understood in 
the context of the previous experimental studies on the $\phi$ meson at 
finite density. 
The E325 experiment at KEK has observed a significant excess 
on the lower mass side of the dilepton spectrum of slowly moving 
$\phi$ mesons produced in 12 GeV $p + A$ reactions, and extracted 
a negative mass shift of the $\phi$ of $35 \pm 7\,\mathrm{MeV}$ at 
nuclear matter density \cite{Muto}. 
In view of the results given in this paper, this finding is somewhat puzzling as such a large 
negative mass shift would correspond to values of $\sigma_{sN}$ larger than 100 MeV 
(see Fig. \ref{fig:sigmaN}), which seems to be in contradiction with recent lattice 
data \cite{Young,Babich,Durr,Horsley,Bali,Semke,Freeman,Shanahan,Ohki,Alarcon,Engelhardt,Junnarkar,Jung,Gong,Alexandrou,Lutz,Ren}, 
which suggest that $\sigma_{sN}$ should at least be smaller than 70 MeV. 
Furthermore let us mention here that according to \cite{Lee}, finite momentum effects will 
at the upper boundary of the 
lowest momentum bin of the E325 experiment ($\beta \gamma < 1.25$) 
lead to a further positive mass shift of 3 MeV for the transverse and 7 MeV for the longitudinal component 
of the $\phi$ meson, and hence are not 
expected to help resolve the above discrepancy. On the 
contrary, they even lead to a further increase of the 
mass and therefore presumably will only worsen the situation. 

The E16 experiment, to be performed at the J-PARC facility, will measure the 
$\phi$ meson in nuclear matter with much better statistics than in E325 and 
will thus hopefully provide much more precise information on the modification of the $\phi$ 
meson spectrum \cite{Kawama}. 
First of all, it will certainly be very interesting to see whether the result of the E325 experiment 
can be reproduced and what mass shift value will be extracted from the experimental data. 

Related to the above topic, it is of course important to ask whether the sum rule approach 
could be missing some important effects and thus not be accurate enough to make 
precise statements on the behavior of the $\phi$ meson spectrum at finite density. 
Here, we want to stress once again that the OPE of this channel is relatively well determined because 
all important terms appear at dimension four or lower. 
This, however, of course does not fully exclude the possibility of 
some so-far neglected contributions 
quantitatively modifying our results in some way. 
Such higher order terms include condensates of higher dimension, further 
$\alpha_s$ corrections and 
terms beyond the linear density approximation. 
Among these, the terms of higher order in $\rho$ are presumably the most 
dangerous ones, as it is for instance known from in-medium chiral perturbation theory that 
the light quark condensate and finite density deviates about $5-7\,\%$ from the linear behavior 
\cite{Kaiser,Goda}. Such a deviation could also exist for the strange quark condensate, 
which would modify our results accordingly. 
Therefore, for making our conclusions more solid, it 
would be desirable to take such kinds of contributions into account. We 
are planing to tackle this task in a future publication. 

\section{\label{Conclusion} Conclusion}
We have studied the behavior of the $\phi$ meson in cold (T=0) matter 
of finite baryonic density. 
This has been done with the help of a QCD sum rule approach, in which the finite density 
effects can be treated as modifications of the condensates of the QCD ground 
state, and thus provides a relation between the conversion of the $\phi$ meson spectrum 
and the change of the various order parameters of QCD. 
We have pointed out that, in the case of the $\phi$ meson, 
there is a strong correlation between the mass shift and 
the strange sigma term $\sigma_{sN}$ 
(shown in Fig. \ref{fig:sigmaN}). 
This correlation 
emerges because of the specific properties of the $\phi$ meson channel. First, it is known that 
the $\phi$ meson remains relatively narrow even at nuclear matter density \cite{Polyanskiy} and 
its broadening thus does not introduce a large uncertainty into the calculation. 
We have checked this point explicitly using a mock data MEM analysis and found that this 
uncertainty is only as large as a few MeV in terms of the $\phi$ meson mass shift 
(see Fig. \ref{fig:mock}). 
Second, the density dependence of the relevant terms of the OPE is well understood, 
the only unknown parameter being $\sigma_{sN}$, and the uncertainty due to 
not well constrained higher order contributions (such as the four-quark condensates) 
is sufficiently small. 

Some more work still remains to be done in the future. 
As a first point, it would be important to test the stability of our results by including 
more higher order terms, especially those that go beyond leading order in density. 
It might also be possible to improve the MEM extraction of the spectral function 
with a recently proposed formulation of QCD sum rules on the complex Borel plane \cite{Araki}. 
Furthermore, it would be interesting to confirm the earlier works on the $\phi$ spectrum at finite momentum 
\cite{Lee,Leupold3}, to study the constraints provided by the sum rules on the finite momentum 
spectrum in detail, 
and to make predictions for the experimental measurements planned to be 
performed in the E16 experiment at J-PARC, at which the $\phi$ spectrum at both 
zero and finite three-momentum will be measured \cite{Kawama}. 

\section*{Acknowledgements}
The authors thank Dr. S. Yokkaichi for motivating us to start 
this work and Professor M. Oka for enlightening discussions on the issues discussed in 
this paper. K.O. gratefully acknowledges the support by the 
Japan Society for the Promotion of Science for Young Scientists (Contract No. 25.6520). 
This work was supported by RIKEN Foreign Postdoctoral
Researcher Program and the RIKEN iTHES project.

\end{document}